\documentstyle[aps,preprint,epsf]{revtex}

\begin{document}
\title{Non-linear growth of periodic patterns}
\author{Simon Villain-Guillot}
\address{Centre de Physique Mol\'{e}culaire Optique et Hertzienne,\\
Universit\'{e} Bordeaux I,\\
33406 Talence Cedex, France}
\author{Christophe Josserand}
\address{Laboratoire de Mod\'{e}lisation en M\'{e}canique, UMR CNRS 7607\\
Universit\'{e} Pierre et Marie Curie, 8 rue du Capitaine Scott, 75015 Paris,%
\\
France }
\maketitle

\begin{abstract}
We study the growth of a periodic pattern in one dimension for a model of
spinodal decomposition, the Cahn-Hilliard equation. 
We particularly focus on the intermediate region, where the non-linearity
cannot be negected anymore, and before the coalescence dominates.
The dynamics is captured through the standard technique of a solubility
condition performed over a particular family of quasi-static solutions.
The main result is that the dynamics along this particular class of solutions
can be expressed in terms of a simple ordinary differential equation.
The density profile of the stationary regime found at the end of the
non-linear growth is also well characterized. Numerical simulations
correspond satisfactorily to the analytical results through three
different methods and asymptotic dynamics are well recovered,
even far from the region where the approximations hold.
\end{abstract}

Pacs numbers~: 05.45.Yv, 47.20.Ky, 47.54.+r \newpage

\section{Introduction}

When a homogenous system departs suddenly from equilibrium, the 
fluctuations around the initial ground state are linearly amplified and for example 
the homogenous phase separates spontaneously into two different more stable 
states. The interfaces which delimit the domains of each phase form a
complex pattern and interact with each other, giving rise to interface
dynamics or pattern formation. Its results can be a slow 
process of coarsening which ends up with only two well separated 
domains. This process of first order phase transition arises particularly for
binary mixtures\cite{wagner} or alloys\cite{hillert}, vapor 
condensation\cite{beg}, ferromagnetic Ising model\cite{halperin} or thin 
films of copolymers \cite{copol}.\\

For the most general, first order transitions initiate in two different ways: 
first, a nucleation process, where the homogenous state is put suddenly in a 
metastable configuration, and an energy barrier has to be crossed before 
the transition appears. This is the typical dynamics of cavitation, for 
instance\cite{bali}. The other method is spinodal 
decomposition where the system leads in a linearly unstable configuration;
such is the situation that we will study here. In this latter case, 
three different regimes are identified in the dynamics: first the linear
instability of the homogenous phase develops from the fluctuations,
leading to the creation of a modulation of the order parameter at a well
defined length scale. The modulations grow
exponentially with time as long as the non-linearities are negligeable.
This stage is very short and results mainly in the selection of a particular
length scale for the process. Non-linearities rapidly slow down the growth of
the modulation resulting in an interface pattern composed of well defined
interfaces delimiting domains containing one of the two stable phases. Remarkably, 
this intermediate stage conserves quite perfectly the modulation width,
so that the resulting pattern is of almost the same length scale than the
one selected initially. Finally, a slow, self-inhibiting dynamics dominates
the last stage of the process, due to the interactions between 
the interfaces. The different regions of each phase coalesce in 
the so-called Ostwald ripening where the number of domains diminishes
whereas their typical size increases. The asymptotic state
is decomposed into two domains, one for each phase. This 
coarsening dynamics is in fact present already from the beginning of the
spinodal decomposition; however, as we will discuss below, its influence
on the two first stages can be often neglected.\\

Hillert\cite{hillert}, 
Cahn and Hilliard\cite{CH} have proposed a model equation for a 
scalar order parameter describing the segregation for a binary mixture.
This model, known as the Cahn-Hilliard equation (C-H later on), 
belongs to the Model B class in Hohenberg and Halperin's 
classification\cite{halperin}. Indeed, different models of phase separation 
have been proposed, depending whether the order parameter is a scalar or a 
vector, or wether it is or is not a conserved quantity (for a review see 
\cite{halperin,cross,gunton}). The (C-H) equation is in fact 
a standard model for phase transition with conserved quantities and
has applications to phase transition in liquid 
crystals\cite{coullet}, segregation of granular mixtures in a rotating 
drum\cite{puri}, or formation of sand ripples\cite{melo,stegner}. It is a
partial differential equation to which a conservative noise is added to
account for thermal fluctuations\cite{cook},  .\\

\begin{figure}[h]
\centerline{ \epsfxsize=8truecm \epsfbox{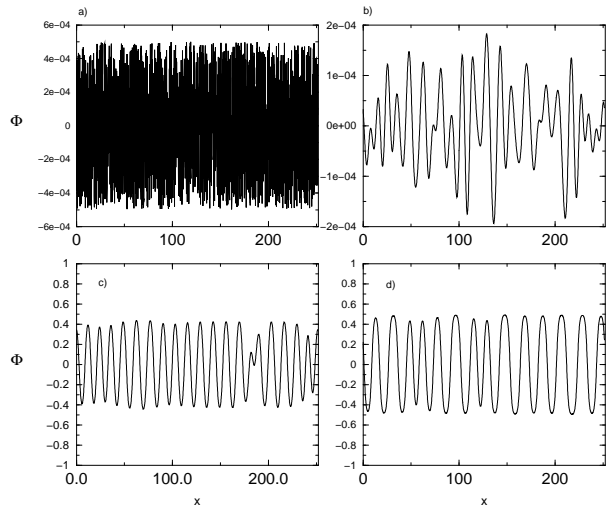}  }
\caption{{\protect\small Time evolution of the order parameter $\Phi (x,t)$
for $\protect\varepsilon =-1$, $dx=0.1227$. (a) initial conditions at $t=0$
are taken randomly with a very low amplitude ($5\cdot 10^{-4}$)~; (b) at
time $t=15$, the amplitude of the modulation has decreased, while only long
wavelength contributions are still present. The small scale perturbations
have been damped by the (C-H) dynamics~; (c) at $t=225$, the modulation has
almost reached its final amplitude, keeping roughly the same number of peaks
as before~; (d) at $t=1800$, we observe that the number of domains has 
decreased from the coarsening dynamics.}}
\label{dyna}
\end{figure}

Figure (\ref{dyna}) shows snapshots of the numerical simulation of the (C-H) dynamics
which represents the full phase transition process after a quench in temperature.
In that case, thermal
fluctuations have been omitted in the dynamics, but were present in the
initial conditions. The three main stages of the spinodal decomposition 
described above are clearly distinguished: first, from Fig. (\ref{dyna} (a)) 
to Fig. (\ref{dyna} (b)), we observe the selection of a typical length scale
for the modulations, then the non-linear growth and its saturation from 
Fig. (\ref{dyna} (b)) to Fig. (\ref{dyna} (c)). We note that the number of 
peaks has been almost conserved between these two configuration; on the
other hand the amplitude of the modulation has now reached almost its
asymptotic value and will not change significantly in the further
dynamics. On the contrary, the coarsening dynamics is observed between 
Fig. (\ref{dyna} (c)) and Fig. (\ref{dyna} (d)) and the typical length of
the pattern is increasing.

Note that in the numerical experiments, contrary to a realistic experience,
we go instantaneously from one temperature to an other (ideal quench).
And we will see in the following that the system will evolve in a smoother
ways in the sense that, due to finite diffusion process, it will not
instantaneously reach the thermodynamical state associated with the temperature
of the quench.

In this paper, we will focus on the one dimensional (C-H) equation and
our aim is to offer a consistent description for the pattern formation,
corresponding to the intermediate (non-linear) regime.
We will obtain a simple ordinary differential equation describing the
dynamics along a family of quasi-static periodic solutions.
We recover the linear regime for short times, and correctly reproduce
the saturation of the second (non-linear) stage, in the case of small
initial perturbations, in the region close to the critical point
(i.e., for a symmetric mixture).
These results are valid in the limit where thermal fluctuations and
coarsening processes are both neglected. We will discuss these important
assumptions and show how the resulting ordinary equation depends on the
wavelength of the periodic solution.\\

As mentionned above, numerous models for phase transition have been proposed ;
an important activity has been devoted to the description of their
dynamics, using both statistical methods or numerical simulations
(for a review see \cite{bray}). However, these works mainly concentrate
on the late stage of the spinodal decomposition where the coarsening
dynamics dominates and which exhibits ''dynamical scaling''~: the dynamics 
presents a self-similar evolution where time enters only through a
length scale $L(t) $, associated with a typical length of the domains
or the rate of decay
of the inhomogeneities. For instance, scaling arguments and stability criteria
give the law $L(t)\sim t^{1/3}$ for spatial dimensions greater than one and a
logarithmic behavior for one dimension in the case of the (C-H) 
equation\cite{bray}.

In this article, by spinodal decomposition, we refer to the two
first stage only, excluding the coarsening dynamics or the third stage.

Only little is known experimentally of these two regimes of the dynamics:
indeed, they are too brief and therefore very hard to capture.
However, these stages were observed in a recent experiment 
on two-dimensional demixion of copolymer\cite{copol} which motivated
our work since it shows the need for a better understanding of the 
dynamics before the coalescence. While the linearized theory gives a full
understanding of the first stage, the second stage of phase separation,
which concerns the saturation of the growth through the non linearity,
appears to have been less studied. There exists numerical attempts to provide
descriptions of the saturation of the profile up to its stationary regime,
using for example a concentration dependant diffusion coefficient,\cite{novik}
$D=D_{o}\Phi (1-\Phi )$, which leads to a modified Kuramoto-Shivashinski equation and
enables to have a saturation of growth.
Here, on the contrary, we work with a constant diffusion coefficient :
the non-linearity will only come from the usual $\Phi ^{4}$ term of the Landau
free energy.

The paper is organized as followed:
first, we present a brief review of general
properties of phase segregations and on
the (C-H) model, mainly to fix the notation. We will reproduce briefly
the original derivation by Cahn and Hilliard, and we will restrict ourselves to
the one dimensional case.
In part III, we motivate the different assumptions of our calculations. 
Numerical
simulations are used to determine the role of the noise and the influence
of the coarsening in the early dynamics.
Then, in part IV we focus on interfaces~; in particular, we will exhibit a
two parameters family of solutions, specific of the one dimensional case,
the so-called soliton lattice.
Finally in section V, we will make use of the solvability criterion in order
to select the dynamical evolution of the density profile among a selected
``ansatz'' solution. 
Eventually, we compare these results with numerical studies of the full
(C-H) dynamics shown at the end. We conclude with a discussion of possible
extensions of this work.

\section{The Cahn-Hilliard model}

The Cahn-Hilliard theory is a modified diffusion equation~; it is a
continuous model, which reads in its dimensionless form:

\begin{equation}
\frac{\partial \Phi }{\partial t}\left( {\bf r},t\right) ={\bf \nabla}^2
(\frac{%
\varepsilon }{2}\Phi +2\Phi ^{3}-{\bf \nabla}^{2}\Phi ).  
\label{CHeq}
\end{equation}

Here ${\bf r}$ and $t$ represent the position vector and the time, the
vectors being noted with bold fonts. $\Phi $ is the order parameter,
a real number~; for instance, it can correspond to the dimensionless
magnetization in Ising ferromagnet, to the fluctuation of density of a fluid
around its mean value during a phase separation or to the concentration of one of
the components of a binary solution in some region around $\overrightarrow{r}$.
$D$ is the diffusion constant and $\varepsilon $ is the dimensionless
control parameter of the system ; it is often identified to the reduced
temperature ($\varepsilon =\frac{T-T_{c}}{T_c}$ where $T_{c}$ is
the critical temperature of the phase transition). This equation, first 
derived  by Cahn and Hilliard\cite{CH}, has also been retrieved
by Langer\cite{langer} from microscopic considerations. As written, the
(C-H) equation does not account for thermal fluctuations present in the
system. They can be added through a Langevin force, which
integrates in the Fokker-Planck equation for the probability distribution
of $\Phi(t)$\cite{langer}. However, as explained by \cite{CH,cook,gunton}, 
the thermal fluctuations can equivalently be taken in account through a 
random noise term on the r.h.s. of equation (\ref{CHeq}). Thus, the (C-H) 
equation reads, in its more general form: 

\begin{equation}
\frac{\partial \Phi }{\partial t}\left( {\bf r},t\right) ={\bf \nabla}^2
(\frac{%
\varepsilon }{2}\Phi +2\Phi ^{3}-{\bf \nabla}^{2}\Phi + \zeta({\bf r},t)), 
\label{CHno}
\end{equation}

where $\zeta$ is a white noise of norm unity, whose amplitude is
proportional to the square root of the temperature of the system.\\

The (C-H) model is a conservative model for the order parameter $\Phi$.
Indeed, it can be written as:

$$ \frac{\partial \Phi }{\partial t}=-{\bf \nabla}\cdot {\bf j} $$

where ${\bf j}$ is the current associated with $\Phi$. Moreover, this 
current obeys the standard law related to the gradient of a so-called 
chemical potential $\mu$ (${\bf j}=-{\bf \nabla} \mu$).
For (C-H), $\mu$ is itself defined as the
functional derivative of a free energy $F$, through:

$$ \mu=\frac{\delta F}{\delta \Phi } $$

with $F$ being in that case the usual Landau-Ginzburg density:

$$ F=\frac{1}{2}\left( ({\bf \nabla} \Phi )^{2}+\frac{\varepsilon }{2}\Phi ^{2}+\Phi ^{4} \right) $$

The homogeneous stationary solutions for the noiseless (C-H) equation are
extrema of the effective potential $V(\Phi )=\varepsilon \Phi ^{2}+\Phi ^{4}$.
For positive $\varepsilon $, there is only one homogenous solution $\Phi =0$
which is linearly stable~; for negative $\varepsilon $, the stationary solution
$\Phi=0$ undergoes a pitchfork bifurcation and three stationary solutions exist.
$\Phi=0$ is still a stationary solution, but it is now linearly unstable ;
two other symmetric solutions $\Phi =\pm \frac{\sqrt{-\varepsilon }}{2}$
are stable and have the same free energy $F=-\varepsilon ^{2}/32$.

Thus, a first order transition can be experienced by quenching the system
suddenly from a positive reduced temperature $\varepsilon$ to a negative
one. Spinodal decomposition is the resulting dynamics. Since for all positive
$\varepsilon$ the system is described by $\Phi=0$,
we only have to study the case where we start at $t=0$ with $\Phi=0$ and a 
negative $\varepsilon$. This is what was shown on figure (\ref{dyna}) in one
space dimension, where the noise has been omitted except for the initial 
condition, where it consists of a random noise of small amplitude around
the mean value $<\Phi >=0$. This can be justified since the noise level,
being proportional to the square root of the temperature, is higher before
the quench than after. Thus, taking a noisy initial condition and omitting
the noise further on can be interpreted as neglecting the noise of the
quenched system compared to the residual noise coming from the ``hot''
initial temperature. However we will discuss more precisely below
the influence of the noise in the quenched phase.\\

When the equation is studied for a constant $\varepsilon$, via a rescaling
of $\Phi$ (as $\sqrt{-\varepsilon}\Phi$), position ${\bf r}$
(as ${\bf r}/\sqrt{-\varepsilon}$) and time (as
$t/{\left| \varepsilon \right|}^2$), we observe that we could restrict 
the dynamics to the case $\varepsilon=-1$. However, since we will later on
compare stationary solutions of (C-H) with different reduced temperature,
we will continue to write the equation with a given $\varepsilon$, keeping in
mind that the dynamics can always be rescaled to the case $\varepsilon=-1$.\\

The stability of the solution $\Phi=0$ can be studied by linearizing
equation (\ref{CHeq}) around $\Phi=0$  (i.e. neglecting the non
linear term $\Phi ^{4}$); considering $\Phi$ as a sum of Fourier modes:

$$ \Phi ({\bf r},t)=\sum_{{\bf q}}\phi _{{\bf q}}e^{i{\bf q\cdot r}+\sigma t} $$

where $\phi _{q}$ is the Fourier coefficient at $t=0$, we obtain that the
amplification factor $\sigma (q)$ satisfies:

$$ \sigma ({\bf q})=-(q^{2}+\frac{\varepsilon }{2})q^{2} $$

\begin{figure}[h]
\centerline{ \epsfxsize=8truecm \epsfbox{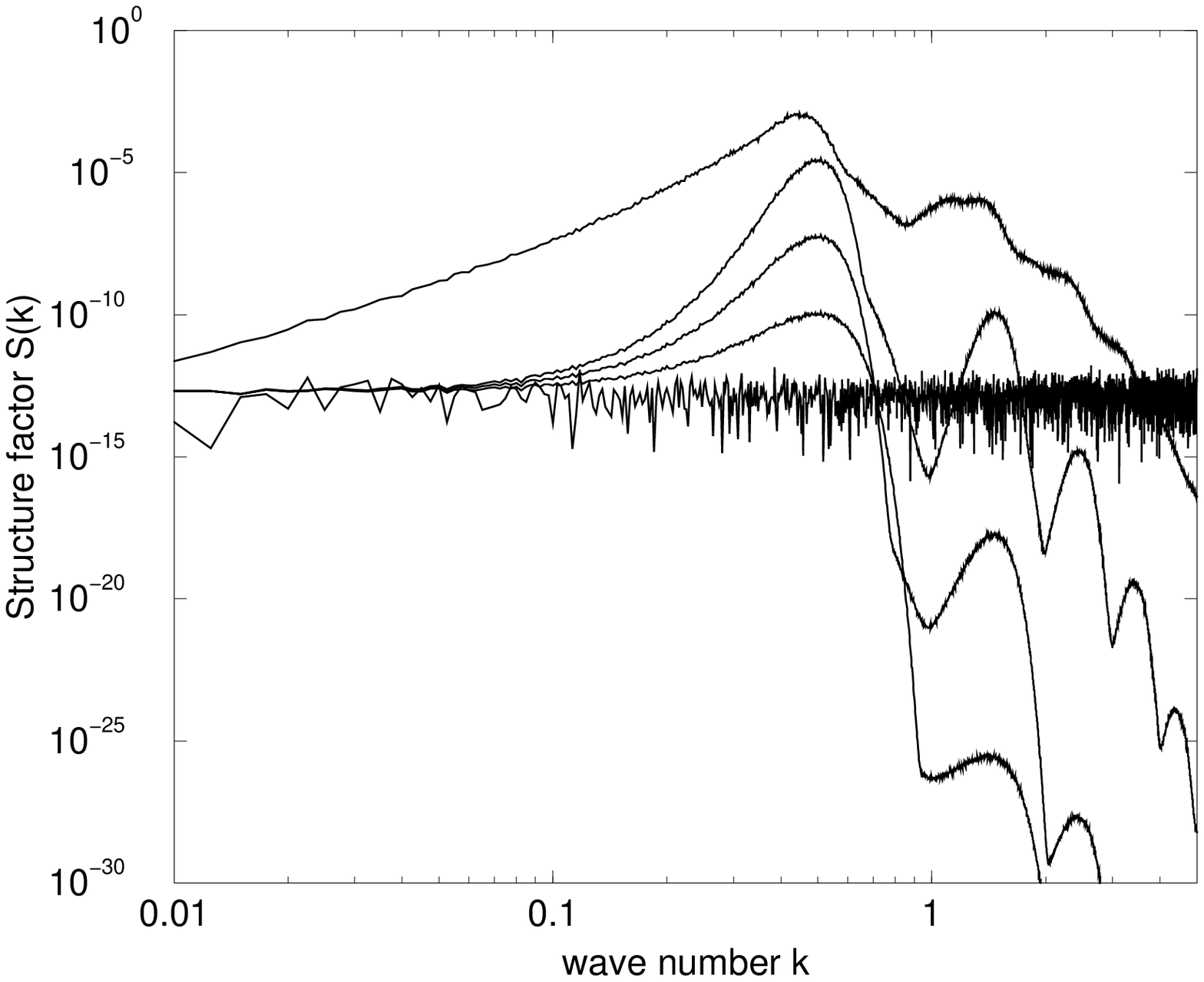}  }
\caption{\protect\small The structure factor $S(k)$ as a function
of the wave number $k$ for different time $t=0,50,100,150$ and $t=500$ 
time units; the higher the peaks are, the larger the time. 
The curves are an average over $100$ initial conditions
taken as a random noise of amplitude $5\cdot 10^{-4}$, the discretization
is over $4096$ grid points with the grid space $dx=0.6$, and $\varepsilon=-1$.
At $t=0$, we observe the flat spectrum of the white noise. For $t=0.5$, the
spectrum reflects the amplification fator: the peaks of the factor is
located at $q_m=0.5$ while for all the modes $q>\sqrt{2}$ the initial
noise has been damped. Then for $t=100$ and $t=150$ we observe the formation
of higher harmonics but the peaks of the structure factor stay around
$q_m$. However, at $t=500.$ the coarsening of the solution has began since
the maximum of $S(k)$ is now at a larger wave length.}
\label{struc}
\end{figure}

It shows immediately that $\Phi=0$ is linearly stable for $\varepsilon >0$
while a band of Fourier modes are unstable for negative $\varepsilon$,
since $\sigma({\bf q}) >0$ for $0<q<\sqrt{(-\varepsilon/2)}$. Moreover, the 
most unstable mode (where $\sigma $ is maximal) is for 
$q_m=\sqrt{-\varepsilon}/2$(with $\sigma _m=\frac{\varepsilon ^{2}}{16}$).
We can anticipate that this wave number of maximum amplification 
factor will dominate the first stage of the dynamics; in particular, it
explains why the modulations appear at length scales close to
$\lambda_m=2 \pi/q_m$, the wave length associated with $q_m$.
Indeed, we show on figure (\ref{struc}) the 
time evolution of the usual structure factor in one dimension:

$$ S(q)=\hat{\Phi }(q) \hat{\Phi}(q)^{\ast } $$

where $\hat{\Phi}$ is the Fourier transform of the field $\Phi$ 
($\hat{\Phi}^\ast$ standing for its complex conjugate). We
have taken the noiseless (C-H) equation with random initial conditions;
the curve is obtained through an average  over $100$ initial conditions.

The different regimes are again well identified: at short times we see
that the modulations whose wave number is close to $q_m$ grow
rapidly from the white noise, while the fluctuations for $q>\sqrt{2}q_m$
for which the amplification factor is negative are damped. Then, 
higher wave numbers emerge, which correspond roughly to harmonics 
mode of the initial modulations. It corresponds to the 
intermediate stage of the
dynamics, where the single mode approximation of the profiles is not valid
any more and the dynamics is in a highly non-linear regime.
 Notice however that the structure factor keeps its peak
located around $q_m$; as we will discuss below, it is indicating that
the number of domains stays almost unchanged during this regime. Later on,
interfaces separating each domain are formed and interact only
through coalescence dynamics: $S(k)$ changes slowly through a
self-similar process (see \cite{langer}) and the peak of the function will
slowly move to smaller wave numbers.\\

\section{An Adiabatic Ansatz}

\begin{figure}[h]
\centerline{ \epsfxsize=8truecm \epsfbox{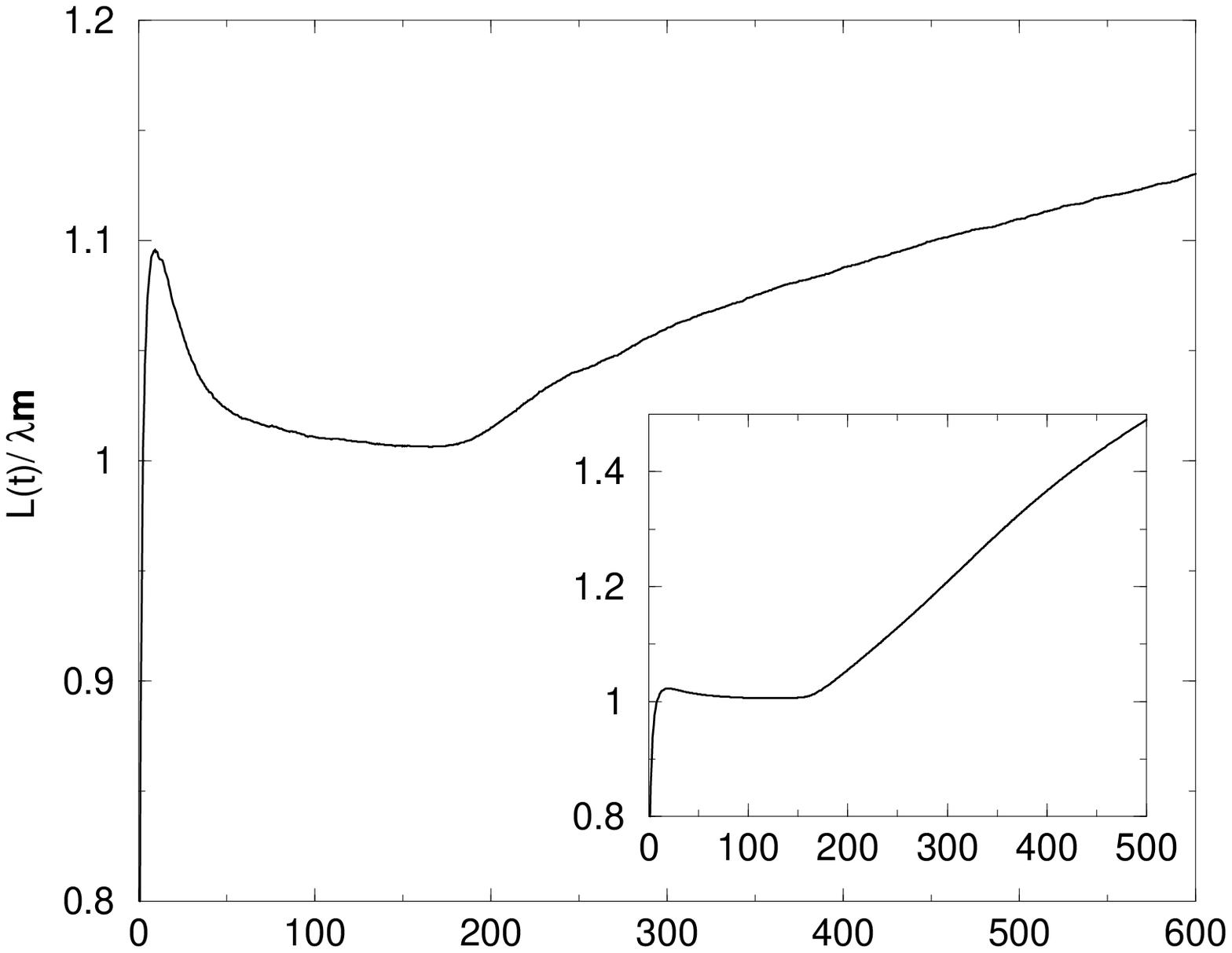}  }
\caption{{\protect\small The ratio between the mean length of the 
modulations and the most amplified wavelength $\lambda_m=2\pi/q_m$ as
a function of time, for the same conditions than figure (\ref{struc}).
The inset shows a similar curve in two space dimensions, obtained by computing
the mean wave number as a function of time, over ten initial conditions.
}}
\label{nint}
\end{figure}

Our analytic method will rely on the assumption that the intermediate region
is approximated through the growth of a periodic modulation solution
of the noiseless (C-H) equation: we need therefore to discuss how this
approach is relevant to the general case where noise is present
and where the coarsening of the non-periodic pattern acts.
Indeed, as it can be seen on figure (\ref{struc}),
the coalescence, roughly characterized by the evolution of the position of
the peak of the structure function, does not appear to influence the
dynamics before a few hundreds of units time. At those times, the intermediate
regime has ended and the modulated pattern is formed. More precisely, 
figure (\ref{nint}) shows the typical mean width of the pattern as a function
of time for the same conditions as figure (\ref{struc}); after a transient 
behavior (until about $t=50$) where the size of the pattern is dominated
by the initial conditions combined with the linear theory of (C-H), we 
observe the intermediate regime (for $t$ between $50$ and $200$ roughly).
In particular, for this regime, we note that the average size of the 
modulation is $\lambda_m$, with a deviation of less than one percent 
from the value predicted by the linear theory. It does not mean that each 
modulation has a length scale of $\lambda_m$, but more precisely that the 
distribution of the modulations length is centered around $\lambda_m$,
as can be seen from the structure factor (see figure (\ref{struc})). 
Moreover, it suggests that the growth 
of each modulation is achieved at constant length scale, determined by the
initial linear instability and thus centered around $\lambda_m$.
At $t$ around $200$, the growth of the modulation is saturate (as can be 
seen from figure (\ref{dyna})) and the coalescence dominates the future
dynamics : the length scale of the structures slightly increases with time.
The inset of figure (\ref{nint}) shows equivalently the typical wave length
of the modulations for the (CH) model in two spatial dimensions; it shows 
again the same plateau that is in favor of the pattern growth at a constant 
size.


\begin{figure}[h]
\centerline{ \epsfxsize=8truecm \epsfbox{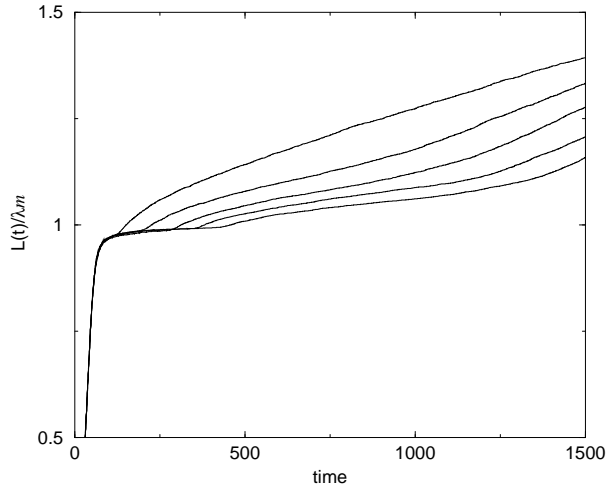}  }
\caption{{\protect\small The ratio between the mean length of the 
modulations and the most amplified wavelength $\lambda_m=2\pi/q_m$ as
a function of time, for different noise level, from $10^{-10}$ to 
$10^{-2}$ (the noise being multiplied by $100$ between each curve).
The shorter the nonlinear plateau, the higher the noise level.
Each curve is obtained through an average over $100$ runs. The other
characteristics of the simulations are the same as for figure (\ref{nint}).
}}
\label{nnoise}
\end{figure}

Thus, we have shown that the coalescence due to the non-periodic pattern selected 
at short time can be neglected during the growth of the modulations.
We need also to quantify the influence of the noise during
the dynamics: until now, we have simplified it to the initial conditions
which induce then a non-periodic initial pattern which is sufficient 
to characterize the general features of the spinodal decomposition.
Moreover, we have shown that the growth of the modulation during the
intermediate regime can be considered to occur at constant length
(centered around $\lambda_m$) for each modulation.
But strictly speaking, noise is always present in the dynamics and,
in addition to feeding the linear instability of the homogenous solution
($\Phi=0$) for short time, it generates a systematic seed of perturbations to
the quasi-periodic pattern. It can therefore disturb this apparent frozen
dynamics at constant size. Figure (\ref{nnoise}) characterizes its effect 
through the evolution of the mean length of the modulation for different 
noise levels. Each curve presents the same behavior, transient dynamics which
selects a length scale of the order of $\lambda_m$, then a plateau regime 
(beginning around $t=50$),
which corresponds to the non linear growth, followed by a coalescence dynamics.
We observe that the length of the plateau regime, on which we are focusing, 
depends strongly on the
noise level.
In particular, for low noise levels, the growth of the modulation seems to 
occur at the constant length $\lambda_m$, during a long period after $t=50$.
But, the higher the noise level, the shorter is this plateau.
The noise stimulates the coalescence process, which
interferes more with the pattern during the intermediate regime\cite{beg}.
For noise levels higher than $10^{-4}$, the growth regime cannot be 
differentiated anymore from the coalescence dynamics and in such cases,
our assumption of growth at a constant scale would no longer be valid.

In conclusion, we can consider that, for low enough noise level, the non
linear growth of the modulations is made at constant length scale and that
the noise has a very weak influence on the dynamics of the two first stages.

Taking advantage of this observation, we can simplify the particular study
of the second stage of the dynamics, the non-linear saturation.
The aim of this paper is therefore to present a
detailed calculation for the growth of a periodic modulation of constant
size, for the noiseless (C-H) equation (\ref{CHeq}) in one spatial 
dimension for a fixed temperature $\varepsilon _{0}$. 

However we hope that this 
approach is valid also in the limit of the small noise levels, where the 
growth of each modulation appears to be unperturbed. 
Although the numerical comparison is developed for the particular case of
$\lambda_m$ periodicity, it applies to any wave length. 

We can now use known results concerning non-homogeneous solutions of the
Ginzburg-Landau equation.

\section{\protect\bigskip Quasi-static approximation}

In order to describe our method, we will first describe
a particular family of stationary solutions of the one dimensional
(C-H).

For $\varepsilon <0$, a stationary solution exists that relies the two
homogenous phases $\Phi =\pm \frac{\sqrt{-\varepsilon }}{2}$

\begin{equation}
\Phi (x)=\frac{\sqrt{\left| \varepsilon \right| }}{2} \tanh (\frac{2x}{\sqrt{%
\left| \varepsilon \right| }}).
\end{equation}

Such a monotonic solution allows a continuum description of the interface
between the two stable phases. In fact, this is a particular member of a one
parameter family of stationary solutions of the Ginzburg-Landau equation

\begin{equation}
\frac{\varepsilon }{2}\Phi +2\Phi ^{3}-\nabla ^{2}\Phi =0
\end{equation}
These solutions, the so-called soliton-lattice solutions~\cite{saxena}, are~: 
\begin{equation}
\Phi _{k,\varepsilon }(x)=k\Delta {\rm Sn}(\frac{x}{\xi },k) \text{ with }%
\xi =\frac{1}{\Delta } =\sqrt{2\frac{k^{2}+1}{-\varepsilon }}
\label{amplitude}
\end{equation}

where ${\rm Sn}(x,k)$ is the Jacobian elliptic function sine-amplitude.
This family of solutions is parametrized by $\varepsilon $ and the modulus
$k\in \left[ 0,1\right] $.
These solutions describe a periodic pattern of period
 
\begin{equation}
\lambda=4K(k)\xi \text{, where }K(k)=
\int_{0}^{\frac{\pi }{2}}\frac{{\rm d}t}{\sqrt{1-k^{2}\sin ^{2}t}}  
\label{period}
\end{equation}

is the complete Jacobian elliptic integral of the first kind.
This family of profiles (or alternating interfaces) can be obtain exactly as
a periodic sum of single solitons and antisolitons (or alternating interfaces) 
\cite{saxena}
\[
\sum_{n}(-1)^{n}\tanh (\pi s(x-n))=\frac{2k(s)K(s)}{\pi s}{\rm Sn}(x,k)%
\hbox{ with  }s=\frac{K(k)}{K(k^{\prime })}\text{ and }k^{\prime 2}=1-k^{2} 
\]

The soliton-lattice solution can be associated with a micro phase separation
locally limited by the finite
diffusion coefficient. For $k=1$, ${\rm Sn}(x,1)=\tanh (x)$, we recover the
usual interface solution~; it is associated with a one soliton solution and
corresponds to a macroscopic segregation. Note that $K(1)$ diverges~; the
solution 
\[
\Phi _{1,\varepsilon }(x)=\frac{\sqrt{\left| \varepsilon \right| }}{2}\tanh (%
\frac{\sqrt{\left| \varepsilon \right| }}{2}x). 
\]
is thus the limit of infinite $s$, when the solitons are far apart one each
others (strong segregation regime).

In the opposite limit ($k\!\rightarrow \!0$, or weak segregation regime)
it describes a sinusoidal modulation
\[
{\rm lim}_{k \rightarrow 0} \Phi _{k,\varepsilon }(x)=k\sqrt{\frac{\left| \varepsilon \right| }{2}}\sin (%
\sqrt{\frac{\left| \varepsilon \right| }{2}}x) 
\]

We now seek the evolution of the solution $\Phi(x,t)$, according to
the (C-H) noiseless dynamics, for a fixed reduced temperature $\varepsilon_0$:

\begin{equation}
\frac{\partial \Phi }{\partial t}\left( x,t\right) =\partial _{xx}(\frac{%
\varepsilon _{0}}{2}\Phi +2\Phi ^{3}-\partial _{xx}\Phi ),  
\label{CH0}
\end{equation}

Numerical simulations of that problem with a small initial condition of
periodicity $\lambda$, shows the growth of the modulation at
this periodicity $\lambda$. As discussed in the previous section,
such
dynamics is unstable and would, in the presence of noise for instance, 
loose its periodicity. However, we have shown that this can be neglected 
for low enough noise levels, and that this ``unstable'' growth of the pattern
is relevant there.

The initial condition will then be taken as the sine mode $q=\frac{2\pi}{\lambda}$ :

$$ \Phi(x,t=0)= \nu \sin(q_{max} x) $$

where $\nu$ is an arbitrary small amplitude. This profile is a member of
the soliton-lattice family (for very small $k$).

The core of the method we are using involvoes to tracking the evolution of
the periodic modulations through a simplified equation. For that purpose,
we now make the {\it ansatz} that at first order, these modulations
belong at any time to the two parameters family of solutions
$\Phi _{k,\varepsilon ^{\ast }}$, with $k$ and $\varepsilon ^{\ast }$
being functions of time. Since the period is chosen to be constant
and equal to $\lambda$, using equations (\ref{amplitude}) and
(\ref{period}), we find that $k$ and $\varepsilon ^{\ast }$ are related
to one another through

\begin{equation}
\varepsilon ^{\ast }(k)= -2(1+k^{2}) \left( \frac{4K}{\lambda}
\right)^2.  
\label{impleps}
\end{equation}

and we have eventually selected a one parameter sub-family of solutions of
given spatial periodicity (that we will call $\Psi ^{\ast }(x,k)$ later on)~:

\[
\Psi ^{\ast }(x,k)=\frac{4K(k)\cdot k}{\lambda}{\rm Sn}(\frac{%
4K(k)x}{\lambda},k). 
\]

\begin{figure}[h]
\centerline{ \epsfxsize=8truecm \epsfbox{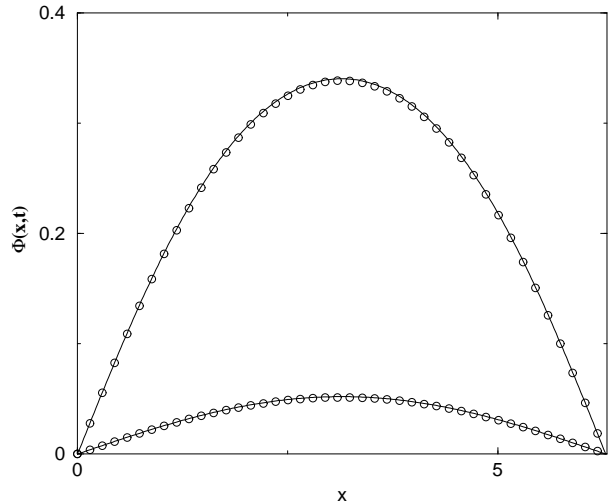}  }
\caption{{\protect\small Comparison for time $t=100$ and $t=140$ between the
numerical solution of (\ref{CH0}) (circles) and the functions $\Psi ^{\ast
}(x,k)$, with $k$ extrapolated from the Fourier transform of $\Phi (x,t)$.
Initial conditions are taken with $\nu =10^{-4}$, and $\lambda=2\pi/q_m$ }}
\label{valide}
\end{figure}

The dynamics of $\Phi (x,t)$ is now reduced to the evolution of $%
k(t)$ (or equivalently $\varepsilon ^{\ast }(t)$). 
Given a function $\Phi $ (obtained
either from experimental data or numerical simulation of equation (\ref{CH0}%
)) at time $t$, the {\it ansatz} assumes that there exists $k$ so that $%
\Phi (x,t)\sim \Psi ^{\ast }(x,k)$. $\varepsilon ^{\ast }(t)$ can be then 
interpreted as a fictitious temperature~:
it is the temperature extracted from the profile at a given time, using
the correspondence between $\varepsilon ^{\ast }$ and $k$ 
of equation (\ref{impleps}). For instance, at $t=0$, the amplitude
is small and we find that $k(0)=\frac{\nu \lambda_m}{2 \pi}$ and thus
$\varepsilon ^{\ast }(0)=8\pi^2/{\lambda}^2$, different {\it a priori}
from $\varepsilon_0$ ($\varepsilon ^{\ast }(0)=\frac{\varepsilon _{0}}{2}$
in the limit $\nu \rightarrow 0$, for $\lambda=\lambda_m$). In the same spirit, we expect that at
the end of the growth,
the ``local temperature'' of the interface coincides with the
thermodynamic one, i.e. the quench temperature $\varepsilon_0$

$$ {\rm lim}_{t \rightarrow \infty} \varepsilon ^{\ast }(t)=\varepsilon_0 $$
at which the dynmics ends.
Somehow, we have assumed 
that the dynamics of (C-H) can be projected at first order onto a
dynamics along the sub-family $\Psi ^{\ast }(x,k)$, which can be 
considered as an attractor of the solutions. This is well
justified when a non sinusoidal initial condition (of small amplitude) is
chosen: we then observe, in numerical simulations, a short 
transient in the dynamics which drives the solution towards the
sine mode at roughly the same amplitude. However, for consistency, we need
to check that at any time, the solution of (C-H) can be well 
approximated by a member of the sub-family. For this purpose, we have developed 
three different algorithms, taking advantage of the general properties of 
the family of
solutions $\Phi _{k,\varepsilon }$ ~: either, $k$ can be deduced both from the
amplitude of the oscillation equals to $ 4kK(k)/{\lambda}$, or from the relation 
$k=\!1-\left( (\Phi ({\lambda}/2,t)/\Phi
({\lambda}/4,t))^{2}-1\right) ^{2}$~; thirdly, a straightforward
computation relates $k$ to the ratio of the two first terms of the Fourier
transform of $\Phi $. We have observed that the three methods show in general
 similar results within an error of one percent. 
However, the validity of the ansatz has still to be
checked by comparing the initial function $\Phi (x,t)$ with the extrapolated
function $\Psi ^{\ast }(x,k)$ obtained by one of these three procedures. It
is shown in figure (\ref{valide}) at two different times in a numerical
simulation of (\ref{CH0})~; we observed that the relative differences
between the two function is much less than $0.01$. 

\begin{figure}[h]
\centerline{ \epsfxsize=8truecm \epsfbox{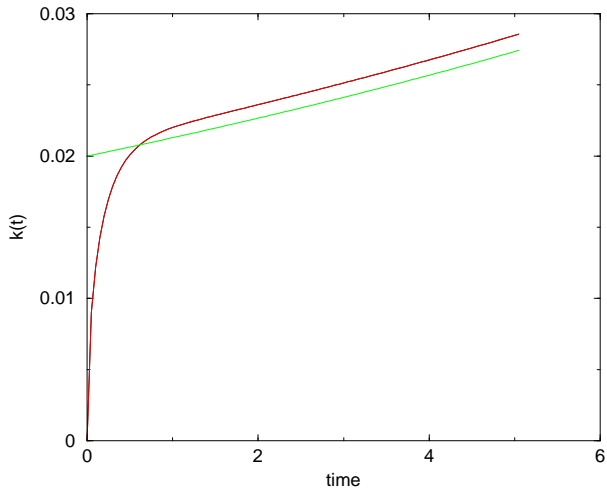}  }
\caption{{\protect\small Comparison for the very early times of the dynamics
between the $k(t)$ predicted by the three different methods. 
The ratio between the Fourier series coefficients and the
ratio between the amplitude at 
$\Phi ({\lambda_m}/2,t)$ and $\Phi({ \lambda_m}/4,t)$ both start at $k=0$.
On the contrary, the third method, which uses the amplitude of the profile,
starts from a finite value.
Nevertheless, the three methods merge within a short time,
indicating the ``affinity''
of the (C-H) dynamics for soliton-lattice solutions.
}}
\label{zoom}
\end{figure}

Moreover, a zoom on the very early time of the one dimensional numerical
simulation with small sinusoidal initial conditions, as presented in
figure (\ref{zoom}),  shows a discrepancy between the
different methods used to extract $k$ from the numerics.
In fact using the ratio between the Fourier series coefficients or the
ratio between the amplitude at two specific points of the profiles,
gives for the very beginning of the numerical simulation the value $k=0$,
in agreement with the statement of the initial condition\footnote{Remark also
 that these two methods give estimations of $k$ that are so close that they 
cannot be differenciated in figure (\ref{zoom}).}.
Indeed, these two methods account for the shape of the profile, which has been 
taken initially as a sine mode, instead of a Sn mode.
	
However, the use the third method, which depends on the amplitude, leads to
a small but finite $k=0.02$ due to the small but finite value of $\nu $.
Even if the distortion from a sinusoidal function associated with this finite
Jacobi modulus is small (the relative change in the natural period and in
the shape of the function is of order $10^{-4}$), one nevertheless observes
that the dynamics of the system is such that, within a short time,
the three methods give results which are again in agreement.
There is a short inflation period during which there is a change in the 
shape of the profile and where
the shoulders of the initial sinusoidal inflate.
That is, there exist a very short stage during which the system goes very
rapidly to a state very close to an element of the family of the soliton lattice
$\Psi ^{\ast }(x,k)$.

\section{Non-linear growth}

Although the evolution of $k(t)$ can be extracted from direct
numerical simulations of (\ref{CH0}), as shown above, the aim of the rest
of this work is to show that $k(t)$ can be deduced via an
explicit ordinary differential equation. Therefore, in what follows, we will
seek the solution of (\ref{CH0}) in the form

\begin{equation}
\Phi (x,t)=\Psi ^{\ast }(x,k(t))+\eta \varphi (x,t)  
\label{ansatz}
\end{equation}

where $\varphi $ accounts for high order correction terms to $\Psi ^{\ast }$, 
while the ``ansatz'' assumes that $\eta \ll \!1$ (we consider $\varepsilon_0$
and $\lambda_m$ of order $1$).

To describe the evolution of the modulus $k(t)$, or equivalently the
dynamics for $\varepsilon ^{\ast }(t)$, we will use the so-called solubility
condition technique. Substituting formula (\ref{ansatz}) in the
Cahn-Hilliard equation (\ref{CH0}), it gives the following dynamics

\[
\frac{\partial \Phi }{\partial t}\left( x,t\right) =\frac{\partial \Psi
^{\ast }}{\partial k}.\frac{{\rm d}k}{{\rm d}t}+\eta \frac{\partial }{%
\partial t}\varphi =\frac{\partial ^{2}}{\partial x^{2}}\left( \frac{%
\varepsilon _{0}}{2}\Psi ^{\ast }+2\Psi ^{\ast 3}-\nabla ^{2}\Psi ^{\ast
}+\eta \left( \frac{\varepsilon _{0}}{2}\varphi +6\Psi ^{\ast 2}\varphi
-\nabla ^{2}\varphi \right) \right) . 
\]
where we have kept only the lowest order terms in the perturbation. As $\Psi
^{\ast }\left( x,k\left( t\right) \right) $ satisfies the relation~:

\[
{\varepsilon ^{\ast }(k)}\Psi ^{\ast }+4\Psi ^{\ast 3}-2\nabla ^{2}\Psi
^{\ast }=0, 
\]
we then have the following dynamics~:

\[
\frac{\partial \Psi ^{\ast }}{\partial k}\!.\;\frac{{\rm d}k}{{\rm d}t}%
+\frac{(\varepsilon ^{\ast }\!-\varepsilon _{0})}{2} 
\frac{\partial ^{2}\Psi ^{\ast }}{\partial x^{2}}+
\eta \frac{\partial }{\partial t}\varphi =\!\eta 
\frac{\partial ^{2}}{\partial x^{2}}(\frac{\varepsilon _{0}}{2}\varphi +
6\Psi^{\ast 2}\varphi -\nabla ^{2}\varphi ) 
\]

The balance of the different terms gives the small parameter of the
expansion $\eta \sim \varepsilon^{\ast } -\varepsilon _{0}$~; we obtain $%
\frac{dk}{dt} \sim \eta$ and $\partial _{t}\varphi \sim \eta \varphi $.
Neglecting the terms of order $\eta^2$ in the previous equation, we end up
solving the linear system~:

\[
\frac{\partial ^{2}}{\partial x^{2}}({\cal L}\varphi )=\frac{\partial \Psi
^{\ast }}{\partial k}.\frac{{\rm d}k}{{\rm d}t}+(\varepsilon ^{\ast
}-\varepsilon _{0})\frac{\partial ^{2}}{\partial x^{2}}\Psi ^{\ast } .
\]

\begin{figure}[h]
\centerline{ \epsfxsize=8truecm \epsfbox{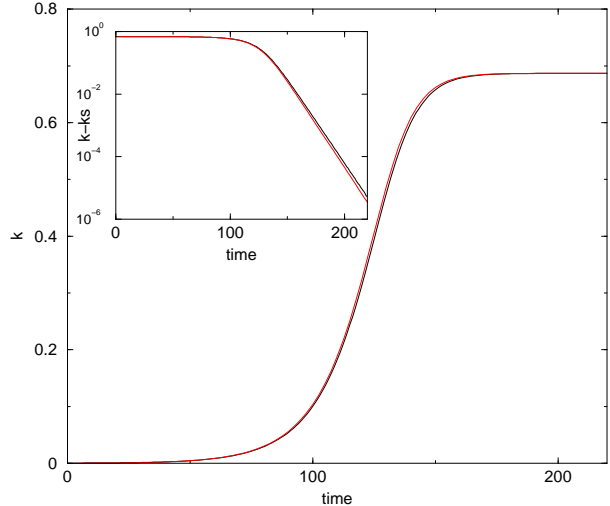}  }
\caption{{\protect\small Comparison between the solution of the ordinary
differential equation (\ref{dynamic}) for the modulus $k(t)$ (gray curve)
and the modulus extracted from the full (C-H) dynamics (black line) with the
same initial condition $k(0)=2\cdot 10^{-4}$, for $\lambda=\lambda_m$.
The dynamics converges to $k_{s}$ in both case for large time.
The inset shows the exponential
convergence of both curves in asymptotic regime, where the solubility
condition is valid~; it compares well with $k_{s}-k(t)\sim e^{-\protect%
\varepsilon _{0}^{2}t/8}$~; in addition, far from $k_{s}$, the exponential
growth for small time $k(t)\sim e^{-\protect\varepsilon _{0}^{2}t/16}$ is
also retrieved by equation (\ref{dynamic}).}}
\label{dieu}
\end{figure}

Here, ${\cal L}$ is the linearized C-H operator ${\cal L}\varphi =\eta \left( 
\frac{\varepsilon ^{\ast }}{2}+6\Psi ^{\ast 2}-\nabla ^{2}\right) \varphi $.
Strictly speaking, this analysis is valid only for $\varepsilon^{\ast } \!\sim
\!\varepsilon _{0}$~; however, it is a classical assumption of the solubility
condition (confirmed below by the numerical results presented on figure (\ref
{dieu})) to expand it for the whole dynamics.\\

A necessary condition for solution is that the right hand side of the system 
is orthogonal to the
kernel of the adjoint operator $\left( \partial _{x^{2}}{\cal L}\right)
^{\dagger }$. The Goldstone mode $\partial _{k}\Phi _{k,\varepsilon ^{\ast
}} $, for $\varepsilon \!=\!\varepsilon ^{\ast }=$const, is clearly an
element of Ker(${\cal L}^{\dagger }$), and if we consider the distribution $%
\chi (x,t)$, such that $\frac{\partial ^{2}}{\partial x^{2}}\chi
(x,t)\!=\Phi _{k,\varepsilon ^{\ast }}(x)$, then we have 
$\partial _{k}\chi \in 
$Ker($\left( \partial_{x^2}{\cal L}\right) ^{\dagger }$)
\footnote{Remark that
the partial derivative with respect to $k$ is made with $\varepsilon ^\ast$ 
taken constant, since we are interested in a member of 
Ker(${\cal L}^{\dagger }$), linearized (C-H) operator for $\varepsilon ^\ast$.}.
Thus, using the
scalar products $<\mid >$ over the period $\lambda $, defined as

\[
<f|g>=\frac{1}{\lambda}\int_{-{\lambda }/2}^{{\lambda }/2}f(x)g(x)dx
\]

we obtain the desired equation for $\frac{{\rm d}k}{{\rm d}t}$~:

\begin{equation}
<\partial _{k}\chi \mid \partial _{k}\Psi ^{\ast }>\frac{{\rm d}k}{{\rm d}t}%
=\frac{(\varepsilon _{0}-\varepsilon ^{\ast })}{2}
<\partial _{k}\chi \mid \partial_{x^{2}}\Phi _{k,\varepsilon ^{\ast }}>=
4(\varepsilon _{0}-\varepsilon ^{\ast})\frac{kK(k)}{{\lambda }^2(1+k^2)}
\left( \frac{2E(k)}{1-k^{2}}-K(k)\right) ,  
\label{diff}
\end{equation}

\noindent where $E(k)=\int_{0}^{\frac{\pi }{2}}\sqrt{1-k^{2}\sin ^{2}x}\,%
{\rm d}x$ is the complete Jacobi elliptic integral of the second kind. The
l.h.s can be expressed using $\psi (x,t)$, defined as $\frac{\partial }{%
\partial x}\psi $=$\Psi ^{\ast }$, which reads~: 
\[
\psi \left( x,t\right) ={\rm ln}\left( {\rm Dn}{(\frac{x}{\xi },k)-k{\rm Cn}(%
\frac{x}{\xi },k)}\right) -{\frac{1}{2}}{\rm ln}{({1-k^{2}})}. 
\]
\noindent ${\rm Cn}$ and ${\rm Dn}$ are the Jacobi elliptic function cosine
and delta amplitudes respectively. Then, noting that

\[
<\partial _{k}\chi \mid \partial _{k}\Psi _{k,\varepsilon ^{\ast
}}>=-<\partial _{k}\psi \mid \partial _{k}\psi >=-I(k) 
\]

where $I(k)$ is independent of $ \lambda $. Finally, equation (\ref{diff})
can be recast as the following explicit ordinary differential equation for
$k(t)$~:

\begin{equation}
\frac{{\rm d}k}{{\rm d}t}=-4
\left( \varepsilon _{0}+2(1+k^2)\left(\frac{4K(k)}{\lambda}
\right)^2 \right)\frac{kK(k)}{{\lambda}^2 I(k) (1+k^2)}
\left( \frac{2E(k)}{1-k^{2}}-K(k)\right)
\label{dynamic}
\end{equation}

In the limit $k \rightarrow 0$, that is, for early times, equation
(\ref{dynamic}) becomes, with the
wave number $q=2\pi /{\lambda}$ associated to the period $\lambda$

$$ \frac{{\rm d}k}{{\rm d}t}=-q^2(\frac{\varepsilon_0}{2}+q^2) \cdot k 
=\sigma(q) k. $$

Since $k$ is proportional to the amplitude of the sine mode of wave number $q$,
we observe that we retrieve the linear theory of (C-H) in that limit.

The r.h.s. of (\ref{dynamic}) is in fact proportional to 
$\varepsilon _{0}\!-\!\varepsilon^{\ast }$, so that the dynamics ends when 
the fictitious temperature reaches the thermodynamic one $\varepsilon _{0}$;
this occurs for $k=k_s$ which satisfies
$32(1+k_s^2)K(k_s)^2=-\varepsilon_0 {\lambda}^2 $.
For $\lambda=\lambda_m$, we obtain $k_s^{2}\!=\!0.471941$.
This corresponds to the 
end of the non-linear growth (on figure (\ref{dyna} c)))
and the value of $k_{s}$ associated with this steady state is well retrieved
numerically by the three methods explained above. Thus, the asymptotic steady 
state solution of equation (\ref{CH0}) for a given period is
${\rm lim}_{t\rightarrow \infty }\Phi (x,t)=\Psi ^{\ast }(x,k_{s})$.
However, no analytic solutions of equation (\ref{dynamic}) have been found, 
and we need to solve it numerically. Figure (\ref{dieu}) compares the 
solution of (\ref{dynamic}) with the dynamic of $k$ extracted by the
Fourier method from the full (C-H) evolution, for the period
$\lambda=\lambda_m$,
the fastest growing mode. It shows a good agreement between the two curves;
in particular, both limit $t \rightarrow 0$ and $t \rightarrow \infty$ are
well captured ; thus equation (\ref{dynamic}) remains valid even for $k$ far
from $k_{s}$.

\section{Conclusion}

We have shown that the choice of an ansatz within the soliton-lattice family
allows a reliable description of the growth of a periodic pattern in the
noiseless Cahn-Hilliard equation. 
Contrary to Ref.\cite{baron}, our ansatz relies on the hypothesis that during
the first two stages of the dynamics, the periodicity of the order parameter
remains constant. In this sense, it is an adiabatic ansatz. The validity of
these assumptions has been investigated in detail
and checked numerically (see Fig.~\ref{valide}). It enables to model the
non-linear growth starting with spatial random initial conditions and
predicts the stationary profile $\Psi _{k_{s}}^{\ast }$,
which ends this non-linear growth.
Although this profile might not be observable in a usual phase transition
due to the presence of noise\cite{langer}, we claim that this approach
should be valid when the noise is low enough, which is the case when the
quench is achieved at low temperatures. We expect this approach to have
a particular pertinence for axial
segregation in rotating drums\cite{oyama}, where the dynamics ends after the
second stage.

The use of the solubility technique combined with the choice of an adiabatic
ansatz might be generalized to the study of other non linear dynamics. For
instance, spinodal decomposition in superfluid Helium or Bose condensate has
been argued to be described by a cubic-quintic non linear equation\cite{rica}%
~; in this particular case, one would first need to retrieve a relevant soliton-like
family of solutions along which to compute the adiabatic dynamics. The same
difficulties would arise when the method is adapted to higher space
dimensions.

Finally, this approach could be used to explore the self-similar scenario
for coalescence suggested by AFM
experiments for spinodal decomposition in mixtures of block co-polymers,
depicted in Ref.\cite{copol}, starting with the previous
stationary distribution as initial conditions. The only change will be in the
use of a family of solutions of growing periodicity $\lambda _{t}$, which would also
be a slow variable of the position, since the coalescence is controlled by
local interactions of the pattern\cite{fraer}. The goal in that case would be
to obtain a differential equation for $\lambda$ the order parameter with the
same technique. However, these questions are postponed to future studies.

The authors are grateful to David Andelman and Sergio Rica for helpful
discussions.

\end{document}